\documentclass{aipproc}
\layoutstyle{8d}
\newcommand{\apj}{ApJ}

\newcommand{\apjl}{ApJL}
\newcommand{\aap}{AAP}
\newcommand{\apss}{ApSS}
\newcommand{\nat}{Nature}
\newcommand{\mnras}{MNRAS}

\SetInternalRegister\hbadness{8000} 

\begin{document}

\title[]{Trans-Relativistic Supernovae, Circumstellar Gamma-Ray
Bursts, and Supernova 1998bw}

\author{Christopher D. Matzner}{
  address={CITA, 60 St George St., Toronto, ON M5S 3H8, Canada},
  email={matzner@cita.utoronto.ca},
  homepage={http://www.cita.utoronto.ca/~matzner},
}

\author{Jonathan C. Tan}{
  address={Astronomy Department, U.C. Berkeley, Berkeley, CA 94720 USA},
  email={jt@astro.berkeley.edu},
}

\author{Christopher F. McKee}{
  address={Depts. of Physics and Astronomy, U.C. Berkeley, Berkeley, CA 94720 USA},
  email={cmckee@astron.berkeley.edu},
}

\copyrightyear{2001}

\begin{abstract}
Supernova (SN) 1998bw and gamma-ray burst (GRB) 980425 offer the first
direct evidence that supernovae are the progenitors of some
GRBs. However, this burst was unusually dim, smooth and soft compared
to other bursts with known afterglows.  Whether it should be
considered a prototype for cosmological GRBs depends largely on
whether the supernova explosion and burst were asymmetrical or can be
modeled as spherical.  We address this question by treating the
acceleration of the supernova shock in the outermost layers of the
stellar envelope, the transition to relativistic flow, and the
subsequent expansion (and further acceleration) of the ejecta into the
surrounding medium.  We find that GRB 980425 could plausibly have been
produced by a collision between the relativistic ejecta from SN 1998bw
and the star's pre-supernova wind; the model requires no significant
asymmetry. This event therefore belongs to a dim subclass of GRBs and
is not a prototype for jet-like cosmological GRBs. 
\end{abstract}
\date{\today}
\maketitle

\section{Introduction}\label{intro}
A growing body of indirect evidence links some long duration gamma-ray
bursts with regions of recent star formation, and therefore with the
core collapse of massive stars (e.g., \citep{kul00}).  The most direct
evidence of such a link is provided by the probable association
\citep{gal98,sof98} of GRB 980425 with SN 1998bw.  However, at the
distance of the supernova, this burst was six orders of magnitude
dimmer than the brightest of cosmological bursts ($10^{48}$ ergs
\citep{gal98}, instead of $10^{54}$, in $\gamma$-ray isotropic
equivalent energy).  Should this burst be considered the first of a
new class of weak, supernova-related GRBs (as proposed by Bloom et
al.\ 1998 \citep{blo98}), or should it be counted among those events
that could produce strong cosmological GRBs?

Central to this question is the degree of asymmetry that must be
invoked to understand the supernova and its GRB.  Evidence of
large-scale asymmetry suggests a jet-like explosion of the core, which
is considered a necessary ingredient for cosmological bursts if they
involve internal shocks within high Lorentz factor flows \citep{sar97}
from the cores of stars \citep{mac99}.  If instead the event is
consistent with spherical symmetry, then it should be inadmissible as
evidence of a causal relation between SNe and any model for GRBs
requiring a jet. 

As a spherical explosion, SN 1998bw possessed about $3\times 10^{52}$
erg of kinetic energy, thirty times more than what is typical of
supernovae.  H\"oflich, Wheeler \& Wang \citep{hof99} argue that SN
1998bw may have been an asymmetric explosion on the basis that this
would allow a lower explosion energy, and Nakamura et al.\
\cite{nak00} find evidence for asymmetry of the inner ejecta in the
late decay of the supernova light curve. Note, however, that the
polarization of light from supernova 1993J suggested asymmetry of its
inner ejecta \citep{1995ApJ...440..821H}, whereas radio emission from
its outermost ejecta \citep{2000Sci...287..112B} shows no asymmetry;
moreover, SN 1998bw exhibited lower polarization than did SN 1993J and most
type II supernovae \cite{1998IAUC.6969....1K}. 

More compelling would be evidence that the observed GRB originated in
a highly asymmetrical event. In this regard,
\cite{1999ApJ...521L.117E}, \cite{1999ApJ...522L.101N}, and
\cite{2001ApJ...546L..29S} advocate a scenario in which a beamed,
highly relativistic outflow is viewed off-axis to produce GRB
980425. 

The competing, more conservative hypothesis construes the burst as the
earliest phase of interaction between high-velocity (spherical)
stellar ejecta and progenitor star's wind -- the same interaction that
gave rise to the later radio emission \citep{li99}.  This possibility
is similar to the suggestion of Colgate \cite{col74} that GRBs might
be due to shock breakout in supernovae.  In this model, the energy
that emerged as gamma rays was previously locked up in the kinetic
energy of expanding ejecta.  Even at the distance of 1998bw, a burst
of GRB 980425's brightness probably required (mildly) relativistic
motion in order to avoid excessive self-opacity.  Corroborating
evidence comes from the very high mean velocity inferred for
the supernova's synchrotron shell ($c/3$ at 12 days; Kulkarni et al.\
\cite{kul98}).

To assess the viability of a spherical model for GRB 980425, we
require: 
\begin{enumerate} 
\item an estimate of the minimum acceptable Lorentz
factor that could have produced the GRB; 
\item an investigation of whether a model for the supernova explosion
that accounts for the optical emission can simultaneously produce
sufficient kinetic energy in material above this Lorentz factor; and
\item a determination of whether the pre-supernova stellar wind was
dense enough to convert the kinetic energy into gamma rays (without
absorbing them) in the duration of the burst. 
\end{enumerate}

\section{Minimum Lorentz Factor of GRB 980425}
The minimum necessary Lorentz factor for GRB 980425 was considered by
Lithwick \& Sari \citep{lit00}, who found it to be at least 3.8 in
order for the burst not be obscured by e$^{+-}$ pairs produced by its
radiation field.  However, this analysis relied on a power law
extrapolation of the observed gamma ray spectrum to energies (in the
comoving frame) above $m_e c^2$.  \cite{lit00} considered spectral
slopes no steeper than -3 ($d\log N_\gamma/d\log e_\gamma$); further,
they adopted $m_e c^2$ as the maximum observed photon energy.  In
contrast, the BATSE light curve for this burst exhibited a 37-$\sigma$
detection in the 50-100 keV channel, 20-$\sigma$ detection in the
$100-300$ keV channel, and no detection at all ($< 1$-$\sigma$) in the
$>300$ keV channel.  These observations give no evidence for photons
with energies above $m_e c^2$.  Interpreted as a power law, the higest
two channels give a slope of -4 or steeper; however, they are more
suggestive of a spectral cutoff (likely a dilute Wien spectrum;
C. Thompson, private communication, 2001) than a power law. 

Another estimate of the minimum Lorentz factor, and one that does not
depend on the specifics of the emission mechanism, arises from the
requirement that $10^{48}$ ergs of gamma rays be produced in an
interaction between stellar ejecta and the pre-supernova stellar
wind. For mean ejecta Lorentz factor $\bar{\Gamma}$, the wind mass
must be about $1/\bar{\Gamma}$ of the ejecta mass. This mass of wind
must be found in a radius that is roughly $2\bar{\Gamma}^2 c$ times
the observed duration of the burst ($\sim 15$ seconds). But, the wind
cannot be opaque at this radius. Applied to the parameters of GRB
980425, these considerations (including the difference between the
velocity of the ejecta and that of the emitting swept-up shell, and
the Klein-Nishina opacity correction) give $\bar{\Gamma}>1.9$,
roughly; see \cite{2001ApJ...551..946T} for a more thorough
discussion.  Both estimates of the minimum Lorentz factor merit
further investigation, preferably careful modeling of both the
dynamical interaction and the emission mechanism; we shall adopt the
latter as the more robust estimate.

\section{Relativistic Ejecta from SN 1998bw} 
As a supernova explosion engulfs a star's envelope, the velocity of
its leading shock front responds to two competing trends: a general
deceleration as increasing mass is swept up, and a tendency to
accelerate down any sharply declining density gradient (in a manner
analogous to the cracking of a whip).  Matzner \& McKee \cite{mat99}
have shown that these trends can be combined into a single formula
that tracks the behavior seen in numerical simulations remarkably
well.  After the shock emerges from the stellar surface, the shocked
material accelerates further as its residual heat is converted into
kinetic energy.  The highest velocity attained by the ejecta is set by
the fact that the shock front spans a finite optical depth; the star
must therefore be relatively compact or have an energetic explosion in
order to produce any relativistic ejecta.  Matzner \& McKee determined
that a compact Wolf-Rayet star would most likely satisfy this
criterion. Although their formulae did not address relativistic
motion, they were able to estimate the kinetic energy in relativistic
ejecta by evaluating their formulae at a final velocity of $c$. This
estimate illustrated that an explosion like that of SN 1998bw would
indeed produce of order $10^{48}$ erg in relativistic ejecta, roughly
enough to power GRB 980425.

Woosley, Eastman \& Schmidt \citep{woo99} considered the production of
relativistic ejecta in the context of specific models developed to fit
the light curve of SN 1998bw. The most promising of these is the
$6.6~M_\odot$ CO core of a $\sim 25 ~M_\odot$ main-sequence star,
exploding with $2.8\times 10^{52}$ ergs of final kinetic energy.
Woosley et al. used the theory of Gnatyk \citep{gna85} to extrapolate
their nonrelativistic simulations into the relativistic regime.  They
concluded that the supernova could not have powered GRB 980425;
however, this conclusion was flawed on several counts. First, Gnatyk's
formula (an interpolation between nonrelativistic \citep{sak60} and
relativistic \citep{joh71} scaling laws) was of unknown
validity. Second, and much more importantly, Woosley et al.\ made an
allowance for the postshock acceleration that was valid in the
nonrelativistic regime (in which the four-velocity $\Gamma\beta$ of a
fluid element increases by a factor 2.5), but did not account for the
very different character of this acceleration found by Johnson \&
McKee \citep{joh71} for relativistic flow (in which $\log[\Gamma
\beta]$ nearly quadruples). This led them to predict a much steeper
decline of kinetic energy with increasing Lorentz factor than actually
holds.  Lastly, Woosley et al.\ assumed that the minimum Lorentz
factor was at least about 5, whereas we have argued above that this
value is not supported by observations and the lower value of $\sim
1.9$ is more appropriate.

To put the theory of this burst on a more solid footing, 
Tan, Matzner \& McKee \citep{2001ApJ...551..946T} have considered in
detail the evolution of explosions involving a transition from
nonrelativistic to relativistic motion. Among the results of this
investigation are: 
\begin{itemize}
\item An extension Matzner \& McKee's analytical theory for the shock
velocity into the relativistic regime, more precisely than in Gnatyk's
theory;
\item Likewise for the postshock acceleration of fluid elements to
their final velocities; 
\item Formulae for the resulting distribution of kinetic energy among
ejecta of different final velocities and Lorentz factors; 
\item An analysis of what aspects of stellar envelopes enhance the
efficiency with which they produce relativistic ejecta; 
\item Simple formulae for the yield of relativistic ejecta from stars
with radiative outer envelopes, in terms of gross parameters
like mass, radius, luminosity and composition; 
\item Formulae to predict the relativistic ejecta in different
directions for numerical simulations of asymmetrical explosions
(including ejecta produced by shock acceleration in beamed and
jet-like events, which could give rise to GRB precursors \citep{mac99});
\item Generalization to the collapses of compact objects (e.g.,
accretion-induced collapse of white dwarfs) in which gravity sets the
characteristic ejecta velocities; and
\item A consideration of the dynamics of putative ``hypernova''
explosions of very high explosion energy. 
\end{itemize}
These analytical results were verified and calibrated by
means of well-resolved, relativistic numerical simulations in
spherical and planar symmetry. 

Tan et al.\ verified Matzner \& McKee's prediction of the kinetic
energy in relativistic ejecta, demonstrating that this energy is
associated with ejecta moving with $\Gamma_f>1.41$.  Applying their
results to Woosley, Eastman \& Schmidt's model CO6 (kindly provided by
Stan Woosley), Tan et al.\ find that the energy of GRB 980425 emerged
in material whose minimum Lorentz factor was 1.7, for which
$\bar{\Gamma}= 2$.  Coupled with the minimum Lorentz factor identified
above, this confirms Matzner \& McKee's prediction that SN 1998bw
produced enough energy in relativistic ejecta to have powered GRB
980425.  For higher Lorentz factors, Tan et al.\ predict a decline in
kinetic energy roughly as $E_k(>\Gamma_f)\propto 1/\Gamma_f$ because
of dramatic postshock acceleration in the relativistic regime. This
relatively shallow decline indicates that explosions that can produce
any relativistic ejecta also channel significant energy into
ultrarelativistic motion. 
\begin{figure}
\caption{Density distribution in the progenitor model CO6 of Woosley,
Eastman \& Schmidt \citep{woo99} (provided by S. Woosley), which was
chosen to match the light curve of SN 1998bw.  Also plotted, for an
explosion of $2.8 \times 10^{52}$ erg, are the four-velocity
($\Gamma\beta$) of the shock front and the final four-velocity in free
expansion, according to the theory of Tan et al.\
\citep{2001ApJ...551..946T}. (This theory sets the final velocity for
the region of terminal shock acceleration; see \cite{mat99} for a
theory of the inner ejecta.) The outermost ejecta with mean velocity
$\bar{\Gamma} \bar{\beta}>1.6$ ($\bar{\Gamma} > 1.9$) may have
contributed to GRB 980425. \label{fig1} }
\includegraphics[height=.5\textheight]{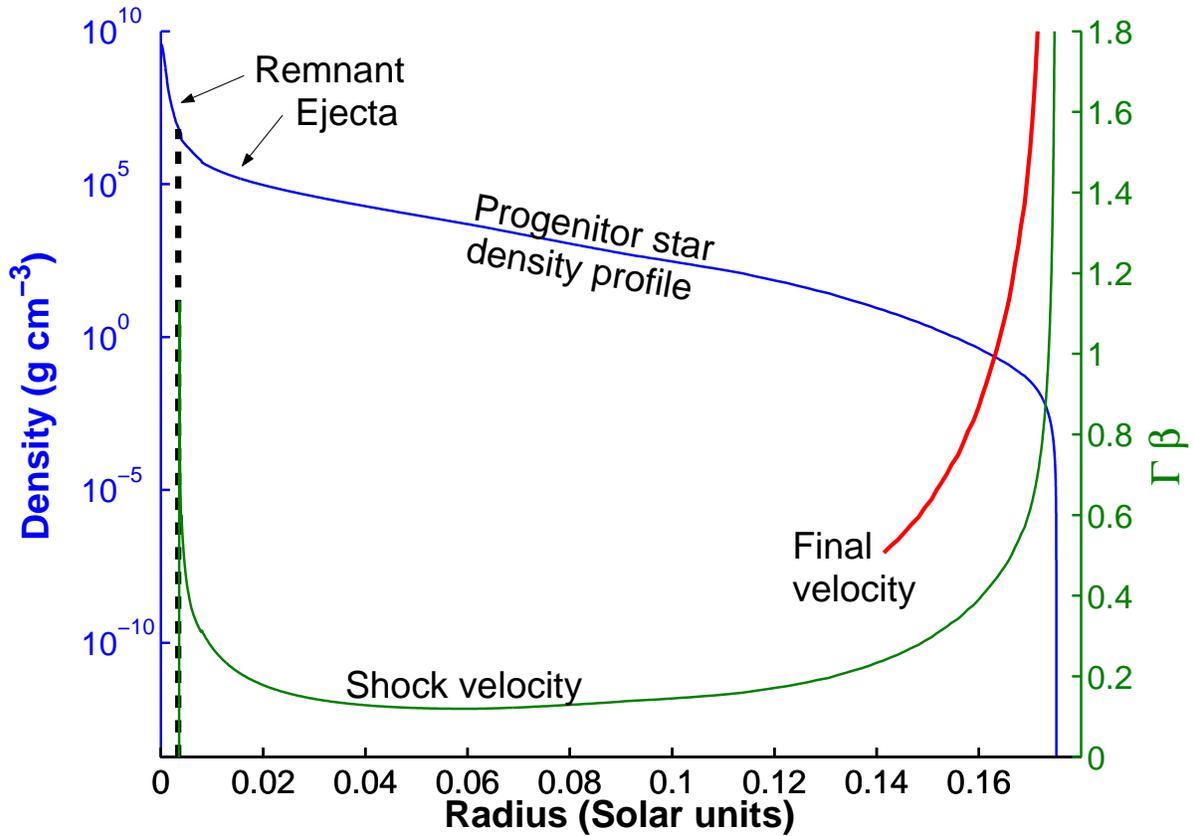}
\end{figure}

\begin{figure}
\caption{Kinetic energy contained in ejecta traveling higher than a
given final velocity, for the explosion depicted in Figure 1,
according to the theory of Tan et al. \citep{2001ApJ...551..946T}. The
observed energy of GRB 980425 is realized in ejecta with final Lorentz
factors above $1.7$. 
\label{fig2} } \includegraphics[height=.5\textheight]{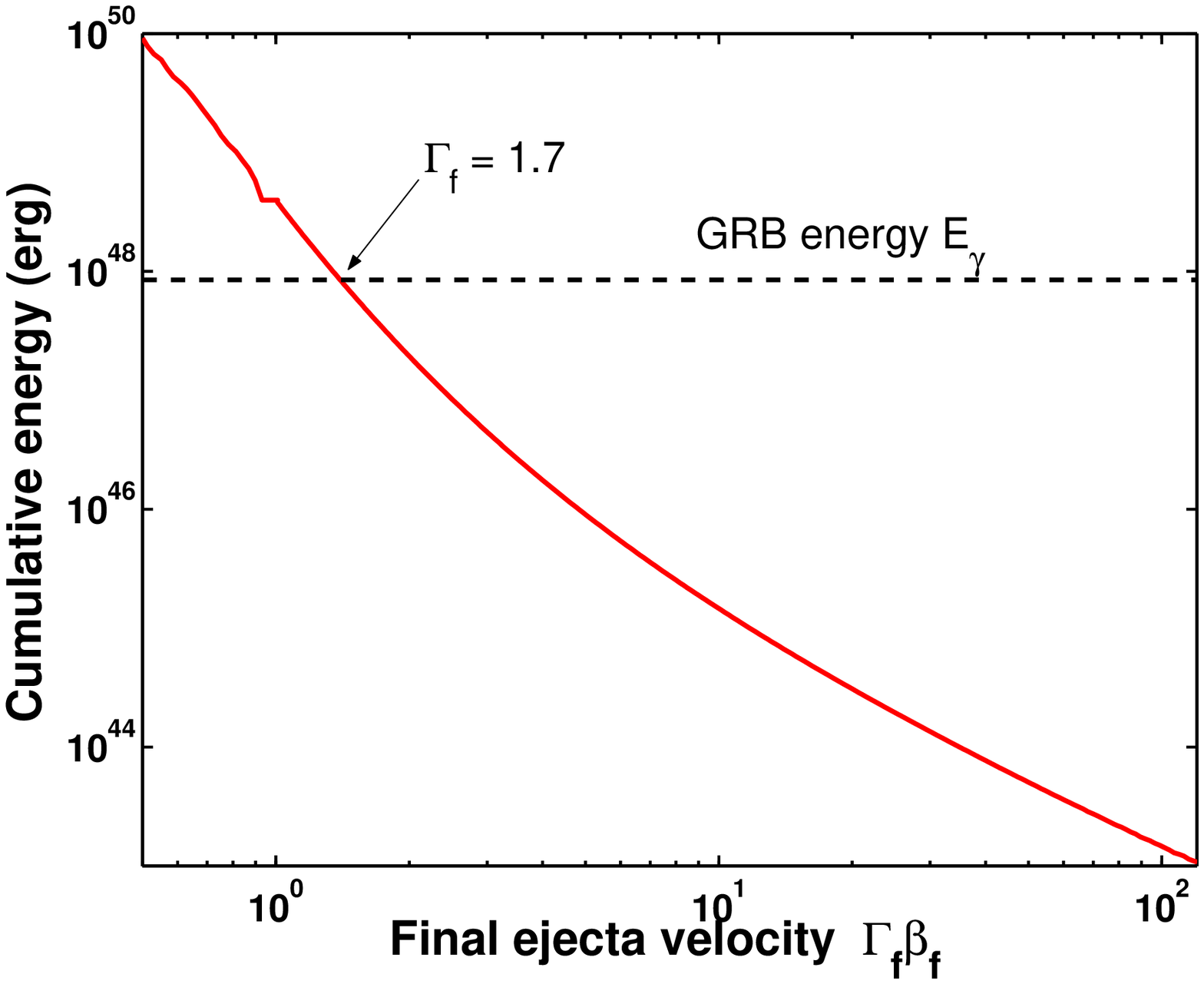}
\end{figure}

\section{Circumstellar Material}
As discussed above, the interaction that gives rise to the GRB occurs
at a radius that is roughly $2\Gamma^2 c t_{\rm obs}$. Within this
radius, a mass $E_\gamma/[c^2\bar{\Gamma}(\bar{\Gamma}-1)]$ of
circumstellar material must be found.  The circumstellar material must
therefore have a mass per unit radius of $\sim E_\gamma/[2 c^3 t_{\rm
obs} \bar{\Gamma}^3 (\bar{\Gamma}-1)]$. The lower limit on
$\bar{\Gamma}$ thus puts an upper limit on the mass per unit length in
the circumstellar material; for a stellar wind, this is the ratio of
mass loss rate to wind velocity. Evaluated for the parameters of
GRB980425, the maximum value of this ratio (attained for the minimum
Lorentz factor) is $(3\times 10^{-4}~M_\odot/{\rm yr})/(1000 ~{\rm
km/s})$. This is dense but within the range of Wolf-Rayet (WC subclass
\citep{1995A&A...299..503K}) winds -- especially considering that the
circumstellar material involved in the GRB was emitted in the last ten
hours of the star's life.

The radio afterglow from SN 1998bw provides a consistency check on any
model in which the GRB arises from an early circumstellar
interaction. Applying the theory of Chevalier
\citep{1982ApJ...258..790C} and Nadyozhin \citep{nad85} to the
collision of the nonrelativistic ejecta with the circumstellar wind,
Tan et al.\ find a mean expansion velocity of $0.35 c$ for the first
12 days of this interaction -- in excellent agreement with the value
$0.3 c$ given by Kulkarni et al.\ \citep{kul98}. Similar agreement is
found with the detailed modeling of the circumstellar interaction by
Li \& Chevalier \citep{li99}. 

\section{Conclusions} 

We have argued that both the gamma-ray burst and the later radio
emission associated with supernova 1998bw can be explained in the
context of a spherical model for its explosion -- the same spherical
model that was proposed by Woosley, Eastman, \& Schmidt \citep{woo99}
to explain its light curve.  The only additional element that must be
included is a relatively dense circumstellar wind, but one within the
range observed around Wolf-Rayet stars.  The viability of a spherical
for GRB 980425 casts significant doubt on the hypothesis that GRB
980425 was intimately related to beamed cosmological bursts.
Specifically, there is no evidence for a jet of high Lorentz factor
material.  

The model we have advocated for GRB 980425 is an external shock model;
Sari \& Piran \citep{sar97} have shown (under the assumption of
relativistic motion) that such models can be ruled out for GRBs
composed of multiple sub-bursts.  However, GRB 980425 exhibited only
one smooth pulse, and is consistent with mildly relativistic motion;
therefore, the external shock model is tenable.

Tan et al.'s analysis demonstrates that the fraction of a supernova's
energy that winds up in relativistic ejecta is enhanced if the 
stellar atmosphere is as diffuse as possible compared to its core.
Stars whose luminosity is comparable to the Eddington limit are ideal
in this regard.  A high explosion energy and low envelope mass are
even more important, as the energy in relativistic motion scales as
$E^{3.6} M_{\rm env}^{-2.6}$. Pre-explosion mass loss therefore enhances
the possibility of a GRB both by increasing the amount of energy in
relativistic ejecta, and by giving rise to the circumstellar wind
necessary for converting this energy into gamma rays in a brief
period. 

Even if subsequent investigations indicate that we have underestimated
the necessary Lorentz factor or total energy of the interaction that
gave rise to GRB 980425, there are two arguments that the hypothesis
of a circumstellar origin should not be abandoned.  First, the kinetic
energy drops only as $1/\Gamma$ to higher Lorentz factors; a higher
minimum $\Gamma$ requires only a proportionally higher total energy in
relativistic ejecta.  Second, this total energy is quite sensitive to
the explosion energy and envelope mass, the degree of central
concentration of the star's atmosphere, and any mild asymmetry that
may have developed in the explosion.  In the model considered, we
identified $8\times 10^{47}$ erg in material expanding with
$\bar{\Gamma}>1.7$; however, small changes to the model would enhance
(or reduce) this yield by factors of several (figure 8 of
\citep{2001ApJ...551..946T}).


\begin{theacknowledgments}
We thank Stan Woosley for kindly providing a model of SN 1998bw's
progenitor star. C. D. M. is supported by an NSERC fellowship; the
research of J. C. T. and C. F. M. is supported by NSF grants AST-
9530480 and AST-0098365.
\end{theacknowledgments}


\end{document}